\documentclass[
aps,
physrev,
reprint,
onecolumn,
amsmath,amssymb,
nofootinbib,
floatfix
]{revtex4-2}

\usepackage{graphicx}
\usepackage{bm}
\usepackage{booktabs}
\usepackage{hyperref}

\hypersetup{colorlinks}

\begin{document}

\title{Bistability in the stalling behavior of thick, cambered airfoils under elevated freestream turbulence}

\author{Jaylon E. McGhee}
\author{John A. N. Farnsworth}
 \email{Contact author: john.farnsworth@colorado.edu}
\affiliation{%
 Ann and H.J. Smead Department of Aerospace Engineering Sciences,\\
 University of Colorado Boulder, Boulder, Colorado 80303, USA
}

\date{TBD.}

\begin{abstract}
The aerodynamic performance of a thick, cambered FFA-W3-301 airfoil section near and beyond stall was examined under elevated freestream turbulence using time-resolved surface pressure measurements for chord-based Reynolds numbers from 125,000 to 275,000. 
Three passive turbulence grids were used to isolate the effects of turbulence intensity and integral length scale, and the results were compared with a clean inflow.
The elevated turbulence largely had a positive effect on the mean performance of the airfoil section, eliminating premature flow separation and recovering lift and drag polars more representative of a higher effective Reynolds number; however, resolving the lift in time showed that the time-averaged polars concealed bistable switching. 
The time-averaged lift was an occupancy-weighted average of two states, and conventional mean and standard-deviation statistics misrepresented the true post-stall unsteady loads.
Beyond the maximum-lift angle of attack, the lift alternated repeatedly between two distinct levels with no characteristic frequency. 
At the highest tested Reynolds number, the switching persisted over an angle-of-attack range of up to nine degrees, more than an order of magnitude wider than that reported for a thin, symmetric section in a low-turbulence inflow.
The lift levels of the two states were also found to be nearly independent of the inflow turbulence level, whereas the fraction of time spent in each differed between turbulence conditions. 
Conditionally averaged surface pressure distributions and proper orthogonal decomposition of the surface pressure consistently showed that neither state is fully attached; the states correspond to two positions of a single separation front on the suction side, a switching accurately captured by a single pressure mode from the decomposition.
\end{abstract}


\maketitle
\section{Introduction}
Thick airfoil sections are typically adopted when structural demands outweigh aerodynamic efficiency, and in these applications the fluctuating loads a section carries can matter as much as its mean aerodynamic performance.
Airfoil performance is characterized almost entirely by time-averaged polars, which assign a single lift and drag coefficient to each angle of attack.
Near and beyond maximum lift, that description can be incomplete.
Even at a fixed angle of attack and freestream velocity, the flow may oscillate at frequencies far below that of vortex shedding~\cite{Zaman1989, Broeren1998} or alternate intermittently between distinct flow states~\cite{Atallah2024}, producing load fluctuations that can be concealed by the time average.
These behaviors have been documented predominantly on thin sections, whose stall originates at or near the leading edge, and for laminar-separation-bubble-induced stall, where the low-frequency unsteadiness has been commonly linked to the growth and bursting of the laminar separation bubble~\cite{Broeren1998}.
Thick sections typically stall from the trailing edge, with the separation point advancing progressively as the angle of attack increases~\cite{McCullough1951}, so their unsteady loading near and beyond maximum lift may differ from that of thinner sections.
In addition to flow-separation-induced unsteadiness, many practical airfoil implementations commonly operate in turbulent inflows, which provide a source of external disturbances absent from most experimental testing in which these behaviors have been identified.

The effects of freestream turbulence on airfoil performance have similarly been characterized mainly through the time-averaged loads.
Elevated turbulence intensity promotes earlier transition of the boundary layer~\cite{Saric2002}, and on airfoils, elevated freestream turbulence consistently delays stall to higher angles of attack.
Its effect on the maximum lift coefficient depends on the geometry, generally with an increase in the maximum lift reported on symmetric sections and a decrease observed on cambered airfoils~\cite{Yap2001,Devinant2002,Kay2020,Damiola2023}.
Studies that vary the turbulence intensity and the integral length scale independently have shown that the mean response is also sensitive to the length scale, particularly when it is of the order of half the chord or larger~\cite{Vita2020,Thompson2023,Cao2011}.
Surface pressure and particle image velocimetry measurements have related these changes to the modification of the laminar separation bubble by the inflow, from the conditions under which the bubble bursts~\cite{Wang2014} to the transition processes that govern its reattachment~\cite{Istvan2018}.
Mishra \textit{et al.}~\cite{Mishra2022} asked whether elevated turbulence could compensate for a low Reynolds number on a wind turbine blade section and found that elevated turbulence brought the mean lift curve much closer to full-scale measurements without matching the lift polar entirely.
They concluded that the unsteady flow, and not only the mean loads, needs to be examined.

The few studies that have examined the unsteady response have shown that turbulence can alter more than just the mean loads.
Li and Hearst~\cite{Li2021} found that intermediate turbulence intensities destabilize the laminar separation bubble and induce low-frequency breathing, and Kay \textit{et al.}~\cite{Kay2020} observed intermittent stall events at a fixed angle of attack and freestream velocity. 
Sicot \textit{et al.}~\cite{Sicot2006} tracked the fluctuating separation point on a 21\%-thick section at turbulence intensities up to 16\% and found the extent of its oscillation was independent of the turbulence level.
That said, they did not determine whether the separation point moved continuously within that range or switched between distinct positions.

Near the maximum-lift angle of attack, the flow over an airfoil does not have to settle into a single state.
Over a finite range of angles of attack, two states with distinct separation topologies and lift levels can coexist, and in polars acquired with increasing (pitch-up) and then decreasing (pitch-down) angle of attack, this coexistence appears as hysteresis, with lift collapsing and recovering at different angles~\cite{Hristov2018}.
From a dynamical-systems perspective, the two branches of such a loop are stable solutions whose ends are marked by saddle-node bifurcations~\cite{Busquet2021}.
When freestream disturbances are sufficient to carry the flow between the different states, the hysteresis loop gives way to intermittent switching.
Atallah \textit{et al.}~\cite{Atallah2024} observed such memoryless switching on a NACA 0012, with transitions triggered by rare, extreme fluctuations; once records were long enough to capture many switching events, no hysteresis was observed in the mean loads between increasing and decreasing angles of attack. 
Boujo \textit{et al.}~\cite{Boujo2025} reproduced these dynamics with a stochastic model whose noise intensity depends on the flow state.
For a 20\%-thick, cambered airfoil section at Reynolds numbers above $10^6$, the wall pressure at the intermittent separation point has been shown to jump between attached and separated levels over an angle-of-attack range of several degrees near the maximum-lift position~\cite{Neunaber2022}. 
These jumps were shown to be anticorrelated between two spanwise stations~\cite{Braud2024}, which a recent preprint attributes to the motion of a stall cell along the span~\cite{Hanna2026}.
In each of these experiments, however, the disturbance level was fixed by the facility, at turbulence intensities of 1.5\% or less.
Atallah \textit{et al.} attributed the switching in their experiments to that background turbulence, and Boujo \textit{et al.} hypothesized that stronger inflow turbulence would alter both the effective forcing and the states themselves~\cite{Atallah2024, Boujo2025}, but no experiment has varied the inflow turbulence to determine whether it changes the states, the switching between them, or both.

The present study examines the stall of a thick, cambered airfoil section under elevated freestream turbulence using time-resolved surface pressure measurements on the FFA-W3-301 airfoil.
A chordwise distribution of pressure ports at midspan was sampled to quantify the variations in the instantaneous sectional lift and pressure drag.
Measurements were acquired for a clean inflow and downstream of three passive grids selected to study the effects of turbulence intensity and integral length scale: two grids produce nearly the same turbulence intensity at different integral length scales, while a third combines a lower intensity with the smallest integral length scale.
These experiments were carried out to address the following research questions.
First, how does freestream turbulence modify the mean lift and pressure drag of a thick airfoil section, and what dependence is there on Reynolds number?
More specifically, is that turbulence-modified response governed primarily by the turbulence intensity, or does the integral length scale also play a dominant role?
Second, does the lift response alternate or jump between two distinct lift states, as has been shown for thinner airfoil sections? 
If so, then over what range of angles of attack do these states persist, and does the inflow turbulence change the states themselves, the fraction of time the flow spends in each state, or both?
Lastly, when investigating the pressure distributions directly, what is the physical source of the variation in the lift and pressure drag observed? 
For thinner airfoil sections, the two-state switching has been linked to dramatic changes in the flow attachment or separation, with the flow jumping between predominantly attached and separated states.
For thicker airfoils such as the one studied here, where separation and stall are known to progress gradually from the trailing edge, does the separation point still jump between states, or does it advance progressively? And where along the chord do these changes occur?

% ============================================================
\section{Methods}
\begin{figure}[t]
  \centering
  \includegraphics[width=5in]{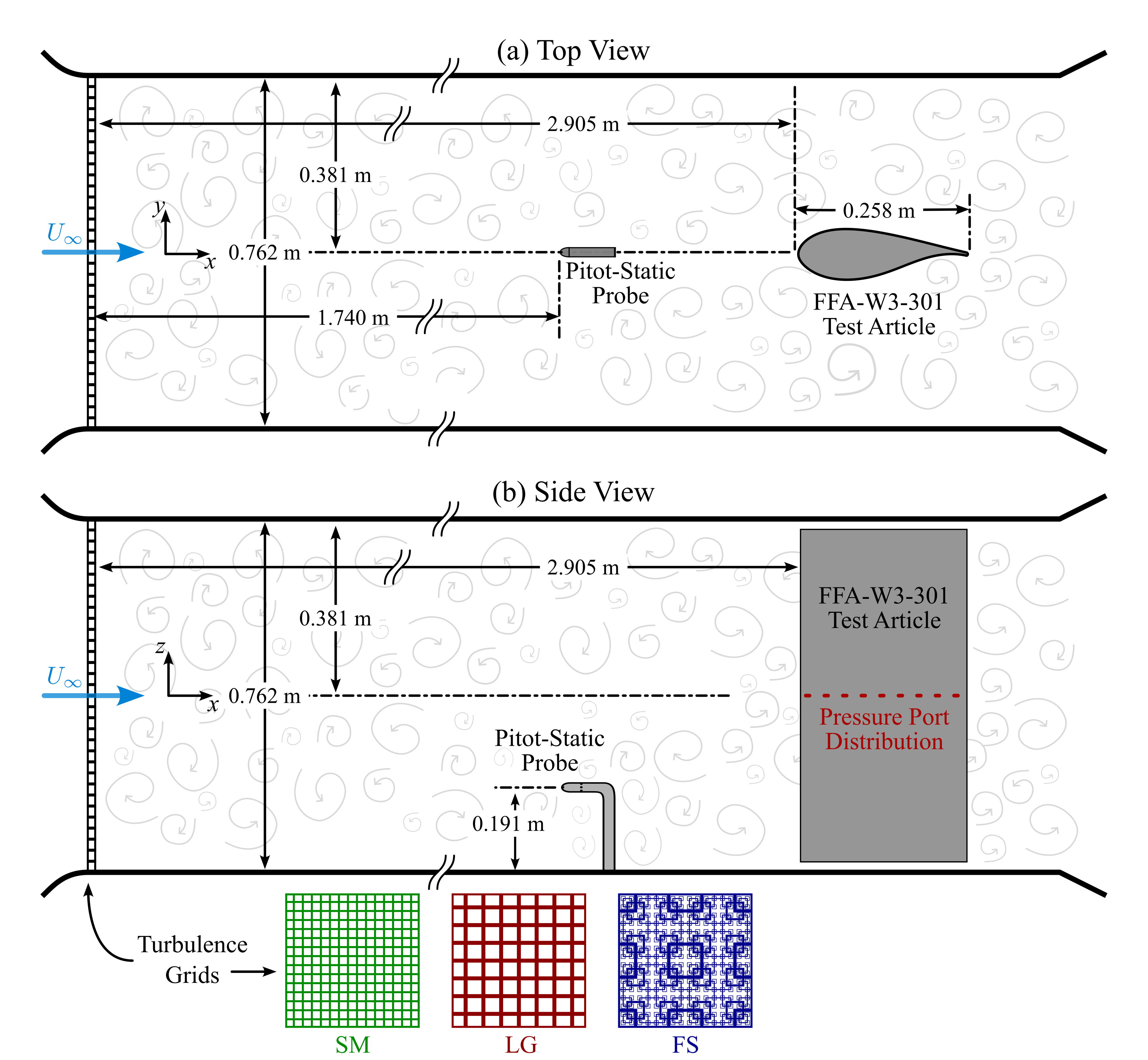}
  \caption{(a) Top view ($x-y$ plane) and (b) side view ($x-z$ plane) of the FFA-W3-301 airfoil test article installed in the wind-tunnel test section. The surface pressure port distribution is located at the midspan of the model. In this study, three passive turbulence grids are installed at the test-section entrance: a small rectangular grid (SM), a large rectangular grid (LG), and a fractal square grid (FS).}
  \label{fig:expstp}
\end{figure}

Experiments were conducted in the low-speed, unsteady wind tunnel at the University of Colorado Boulder~\cite{Farnsworth2020}. 
A schematic of the experimental setup is shown in Fig.~\ref{fig:expstp}. 
The wind tunnel test section is 3.58~m long with a 0.762~m $\times$ 0.762~m cross section.
Three passive turbulence grids were interchangeably installed at the test-section inlet.
Note that these include a small rectangular grid (SM) with a mesh size of 0.052~m, a large rectangular grid (LG) with a mesh size of 0.102~m, and a four-generation fractal square grid (FS) with a width of 0.191~m for the first generation~\cite{Sheppard2025}. 
The FFA-W3-301 airfoil model, which spanned the test-section width, was installed 2.905~m downstream from the grid plane. 
Freestream reference conditions were measured using a Pitot-static probe located 1.740~m downstream from the grid plane, or approximately 4.5 chord lengths upstream of the model, and 0.191~m from the test-section floor to ensure its wake did not interfere with the pressure ports instrumented at the midspan of the airfoil model.

\begin{figure}[t]
  \centering
  \includegraphics[width=3.5in]{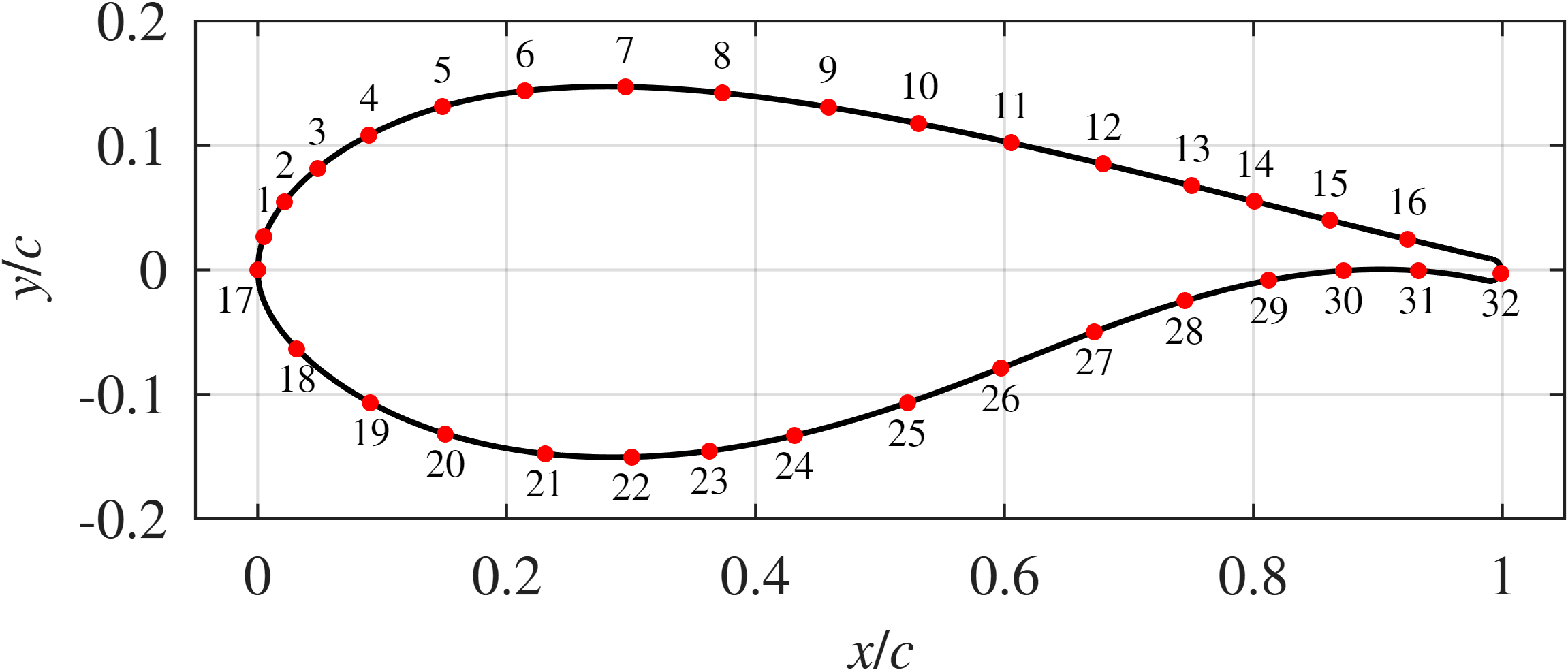}
  \caption{Cross section of the FFA-W3-301 test article ($c = 0.258$~m) displaying the chordwise locations of the 32 surface pressure ports.}
  \label{fig:airfoil}
  \end{figure}

The model was an FFA-W3-301 airfoil, a 30\%-thick section used at the inboard stations of the IEA 15-MW and 22-MW reference wind turbines~\cite{Gaertner2020,Zahle2024}.
The model had a chord length of $c = 0.258$~m and a span of $b = 0.756$~m, leaving 3.2~mm gaps at the test-section floor and ceiling and yielding a span-to-chord ratio of approximately 3.
Thirty-two static pressure ports, each 0.76~mm in diameter, were located at midspan at the chordwise positions shown in Fig.~\ref{fig:airfoil}. 
The ports were connected to a 32-channel Scanivalve ZOC23B pressure scanner through internal passages within the airfoil model, which was printed using stereolithography. 
The model was mounted on a steel shaft at the quarter-chord, and the angle of attack, $\alpha$, was set by a motor system through a planetary gearbox with a commanded resolution of 0.15$^\circ$ and 0.1$^\circ$ of mechanical backlash.

The Scanivalve ZOC23B pressure scanner (range $\pm 2.5$~kPa, accuracy $\pm 5$~Pa) measured the differential pressure at each port relative to a sealed reference volume vented to atmospheric pressure before each run.
The static ports were sampled sequentially at 20~kHz, giving an effective rate of 625~Hz per channel. 
Note that the resulting 1.6~ms skew across a single scan is small compared with the timescales of the state switching examined here and discussed in the results section.
The frequency response of the combined tubing and internal model passages for the pressure measurements was estimated using the model of Bergh and Tijdeman~\cite{BandT1975}, which  predicted a flat amplitude response up to approximately 100~Hz for all of the static ports.
Note that the results of this analysis are not presented here for brevity.

The freestream dynamic pressure $q_\infty$ was measured across the Pitot-static reference probe using a heated differential capacitance manometer (MKS Baratron 220DD, range: 100 Torr, resolution: $0.01\%$ of full scale, accuracy: $\pm 0.15\%$ of reading), and the atmospheric pressure was measured from the static ring on the Pitot-static reference probe using a similar absolute capacitance manometer (MKS Type 220DA Baratron, range: 1000 Torr, resolution: $0.01\%$ of full scale, accuracy: $\pm 0.15\%$ of reading). 
The air density was computed from the atmospheric pressure and the air temperature measured in the contraction using the ideal gas law, corrected for relative humidity, and was found to be $\rho_\infty \approx 1.0~\mathrm{kg/m^{3}}$ at the elevation of the facility during all of the tests.
The dynamic viscosity, $\mu$, was obtained from Sutherland's law with the measured air temperature.
The freestream speed and chord Reynolds number were computed following typical convention: $U_\infty = \sqrt{2q_\infty/\rho_\infty}$ and $Re = \rho_\infty U_\infty c/\mu$ using the conditions obtained for each test case.

\subsection{Turbulence Generation and Inflow Characterization}
The inflow was characterized at the model location, with the model removed, using single-component hot-wire anemometry and planar particle image velocimetry (PIV) at a nominal freestream speed of $U_\infty = 20$~m/s.
Hot-wire measurements were acquired on the test-section centerline and were sampled at 50~kHz, using four 2-minute records per grid.
The PIV field of view was $0.342$~m$~\times~0.213$~m, and velocity fields were computed from 1,000 image pairs, sampled at a slow rate of 50~Hz to provide uncorrelated measurements, using a final interrogation window size of $32 \times 32$~px at 75\% overlap.
The turbulence intensity is defined as $\mathrm{TI} = u'/\overline{U}$, where $u'$ is the root-mean-square of the streamwise velocity fluctuations.
For the hot-wire records, the streamwise integral length scale was computed from the temporal autocorrelation $R_{uu}(\tau)$ of the velocity fluctuations using Taylor's frozen-turbulence hypothesis,
\begin{equation}
L_{uu,x} = \overline{U}\int_{0}^{\tau_0} R_{uu}(\tau)\,d\tau,
\end{equation}
where $\tau_0$ is the first zero crossing of $R_{uu}$.
For the PIV fields, $L_{uu,x}$ and $L_{uu,y}$ were instead obtained by integrating the spatial two-point correlation of the streamwise fluctuations along $x$ and $y$, respectively, to its first zero crossing, so that no convection hypothesis is required.
The resulting turbulence properties are compared in Table~\ref{tab:inflow} for each of the passive turbulence grids.
The two techniques agree to within approximately 5\% in turbulence intensity and 8\% in the streamwise integral length scale, $L_{uu,x}$. 
Throughout this paper, the turbulence statistics obtained from the hot-wire measurements are used due to their larger sample size and finer temporal resolution.
The primary role of the PIV measurements was to assess the spatial homogeneity of the inflow across the region occupied by the airfoil model. 
In this region, the mean velocity was found to vary by less than $\pm 1.5\%$ and the turbulence intensity by less than 0.5\%.
The PIV also provided an evaluation of the turbulence isotropy, where $u'/v'$  ranged between 1.09 and 1.18 and $L_{uu,y}/L_{uu,x}$ ranged  between 0.53 and 0.60, both indicating nearly isotropic inflow turbulence.
Finally, note that the $\overline{U} =20$~m/s condition corresponds to $Re = 275{,}000$ for the airfoil.
Additional measurements at $\overline{U} = 10$ and 15~m/s returned nearly identical values, and the grid-generated turbulence properties are therefore taken to hold across the range of Reynolds numbers tested.

With regard to the choice of grids used in the present study, the two classical rectangular grids were chosen to provide a well understood turbulence structure used regularly throughout the literature~\cite{Roach1987}.
The mesh sizes of the two classical rectangular grids were chosen to provide a streamwise integral length scale less than the airfoil chord length, $c/L_{uu,x} \approx  7$ (SM) and $c/L_{uu,x} \approx  4.2$ (LG), so as to couple more closely with the airfoil boundary layer as compared to the unsteady response of the airfoil itself~\cite{Amitay2002}, while also producing a significant variation in turbulence intensity between the grids: $\mathrm{TI} \approx 2.5$\% (SM) and $\mathrm{TI} \approx 4.0$\% (LG).
The fractal square grid (FS) was designed to match the turbulence intensity of the large rectangular grid (LG), $\mathrm{TI} \approx 4$\%, while also producing a different turbulence structure such that the integral length scale was moderately smaller, $c/L_{uu,x} \approx  6$ (FS).
Extensive details about the development and structure of different types of fractal grids are given by Hurst and Vassilicos~\cite{Hurst2007}, while a detailed discussion on the design and characterization of the grids used in this study is provided by Sheppard~\cite{Sheppard2025}.

\begin{table}[t]
\caption{\label{tab:inflow}%
Summary of the measured inflow turbulence properties for each grid configuration at a nominal freestream velocity of 20~m/s. Inflow properties were acquired at the airfoil location with the model removed.}
\begin{ruledtabular}
\begin{tabular}{lccccccccc}
 & \multicolumn{3}{c}{Hot-wire anemometry} &
 & \multicolumn{5}{c}{Particle image velocimetry} \\
\cline{2-4}\cline{6-10}
Config. & $\overline{U}$ (m/s) & TI (\%) & $L_{uu,x}$ (mm) &
 & $\overline{U}$ (m/s) & TI (\%) & $L_{uu,x}$ (mm) & $L_{uu,y}$ (mm) & $u'/v'$ \\
\colrule
SM & 19.31 & 2.46 & 38.1 & & 20.12 & 2.57 & 36.0 & 21.2 & 1.18 \\
LG & 19.30 & 4.03 & 60.8 & & 20.11 & 4.08 & 55.7 & 33.2 & 1.11 \\
FS & 19.42 & 3.95 & 45.9 & & 19.80 & 4.14 & 45.7 & 24.1 & 1.09 \\
\end{tabular}
\end{ruledtabular}
\end{table}

\subsection{Test Matrix and Data Reduction}
Surface pressures were acquired at four chord-based Reynolds numbers, $Re = 125{,}000$, $175{,}000$, $225{,}000$, and $275{,}000$, for the clean configuration (i.e., no turbulence grid installed) and with each of the three grids, giving sixteen unique inflow and Reynolds number combinations.
At each combination of test conditions, the angle of attack followed a three-segment trajectory, $0^\circ \rightarrow +30.6^\circ \rightarrow -30.6^\circ \rightarrow 0^\circ$, in which the first and third segments form the pitch-up portion of testing and the second forms the pitch-down portion.
The pitch-up portion was run in $1.8^\circ$ increments, and the pitch-down portion was run at the angles offset from these by $0.9^\circ$, so that the two portions together resolve the polar at a resolution of $0.9^\circ$.
After each change in angle, the flow was allowed to settle for approximately 5~s before a 30~s record of surface pressures was acquired by the pressure scanner.

The surface pressure distribution was integrated around the airfoil contour at every time point to obtain the instantaneous sectional loads.
The pressure coefficient at each port is
\begin{equation}
C_p = \frac{p - p_\infty}{q_\infty},
\end{equation}
where $p$ is the measured surface pressure, $p_\infty$ is the freestream static pressure, and $q_\infty$ is the dynamic pressure from the Pitot-static probe.
The normal and axial force coefficients, defined relative to the chord line, were obtained from the closed-contour integrals
\begin{equation}
c_n = -\frac{1}{c}\oint C_p \, dx,\qquad \text{and} \qquad c_a = \frac{1}{c}\oint C_p \, dy,
\end{equation}
with the contour traversed clockwise from the leading edge. 
These were then resolved into the lift and pressure drag coefficients in wind axes using the angle of attack of the airfoil in the standard way,
\begin{equation}
c_l = c_n \cos\alpha - c_a \sin\alpha, \qquad \text{and} \qquad c_{d,p} = c_n \sin\alpha + c_a \cos\alpha.
\end{equation}
The integrals were evaluated using a first-order finite integration (i.e., trapezoidal rule) between adjacent ports, with the contour closed across the trailing edge.
Only the pressure contribution to the force is captured by the surface pressure measurement, so $c_{d,p}$ excludes skin friction, which was not measured independently.
Further note that the omitted skin friction contribution is expected to be small relative to the pressure drag in the stalled regime due to the thickness-to-chord ratio of the airfoil, $t/c = 30.1\%$.
At $Re = 275{,}000$, the 30~s record spans approximately $2{,}300$ convection times based on the chord, $T_c = U_\infty t/c$, and approximately $10{,}000$ to $16{,}000$ based on the streamwise integral length scale of the inflow, $T_{L} = U_\infty t/L_{uu,x}$, depending on the grid.
These records are proportionally shorter for lower Reynolds numbers, with approximately $1{,}000$ chord-based convective times at the minimum Reynolds number of $Re = 125{,}000$.

The force coefficients and angles of attack reported here were corrected for solid blockage, wake blockage, and streamline curvature following Barlow \textit{et al.}~\cite{Barlow1999}.
The frontal blockage of the model is approximately 10\% at $\alpha = 0^\circ$ and grows significantly near stall.
Because the wake-blockage term depends on the drag, it was evaluated iteratively with the corrected drag coefficient.
The correction factors were computed from the mean force coefficients at each angle of attack and applied uniformly to the instantaneous series, so the correction introduces no fluctuation of its own into $c_l(t)$.

The measurement uncertainty is largest at the lowest Reynolds number, where the dynamic pressure is smallest.
The differential pressure scanner accuracy of $\pm 5$~Pa is approximately 13\% of the dynamic pressure at $Re = 125{,}000$ ($q \approx 40$~Pa) but only 2.5\% at $Re = 275{,}000$ ($q \approx 195$~Pa).
Propagating the instrument uncertainties through the data reduction using a Monte Carlo simulation provided a lift-coefficient uncertainty of approximately $\pm 0.01$ at $Re = 275{,}000$ and approximately $\pm 0.045$ at $Re = 125{,}000$, which was found to be nearly independent of angle of attack.

\subsection{Analysis Methods: State Identification and Proper Orthogonal Decomposition}
\label{sec:analysis}
\begin{figure}[t]
  \centering
  \includegraphics[width=4in]{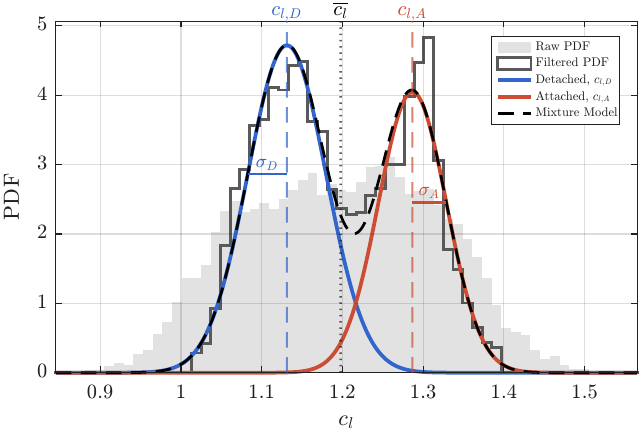}
  \caption{Two-component Gaussian mixture model fit to the 3~Hz low-pass-filtered lift coefficient for the fractal square grid (FS) at $Re = 275{,}000$ and $\alpha = 18.1^\circ$ during the pitch-up portion of testing.
  The histogram of the filtered signal is outlined in black, and the probability density function (PDF) of the unfiltered signal is shown in light gray for reference.
  The lower-lift (blue) and higher-lift (red) components, with means $c_{l,D}$ and $c_{l,A}$, sum to the mixture model (dashed black).
  Horizontal bars mark one standard deviation from each component mean, and the dotted vertical line marks the time-averaged lift coefficient $\overline{c_l}$.}
  \label{fig:gmm}
\end{figure}

Beyond the stall angle, the lift coefficient was found to alternate between two states, so its overall distribution differed from a simple Gaussian model and its resulting statistical mean and standard deviation described neither state alone.
This bimodal nature meant that the states had to be identified from the histogram of the time series directly.
Furthermore, the fluctuations at the timescale of the inflow turbulence worked to broaden the instantaneous distributions of the two states until they overlapped (gray distribution in Fig.~\ref{fig:gmm}).
As a result, prior to identifying the states, the lift coefficient was low-pass filtered using a fourth-order, zero-phase Butterworth filter with a cutoff of 3~Hz, corresponding to a chord-based Strouhal number $St_c = f_c c/U_\infty \approx 0.04$ at $Re = 275{,}000$.
This is approximately two orders of magnitude below the eddy-passage frequency $U_\infty/L_{uu,x} \approx 430$~Hz of the FS grid inflow, so the filter was used to separate the timescale of the switching from that of the turbulence advection.
For comparison, Atallah \textit{et al.}~\cite{Atallah2024} used a 1~s moving average, which suits mean residence times of seconds to hundreds of seconds.
The residence times observed in this study are between approximately 0.5 and 3~s, comparable to that window, so a 1~s average would blend the states rather than isolate them.

The probability density function of the filtered lift coefficient is modeled as a two-component Gaussian mixture,
\begin{equation}
\mathrm{PDF}(c_l) = \frac{\pi_A}{\sigma_A\sqrt{2\pi}} \exp\!\left[-\frac{\left(c_l - c_{l,A}\right)^2}{2\sigma_A^2}\right]
       + \frac{\pi_D}{\sigma_D\sqrt{2\pi}} \exp\!\left[-\frac{\left(c_l - c_{l,D}\right)^2}{2\sigma_D^2}\right],
\end{equation}
where each term describes the distribution of the lift coefficient while the flow occupies one state (Fig.~\ref{fig:gmm}).
The mean $c_{l,i}$ is the lift level of state $i$, the standard deviation $\sigma_i$ is the scatter of the filtered lift coefficient within that state, and the weight $\pi_i$, with $\pi_A + \pi_D = 1$, is the fraction of the record the flow spends in a given state.
The subscripts follow the notation of Atallah \textit{et al.}~\cite{Atallah2024}, with $A$ denoting the component of higher lift and $D$ that of lower lift.
In this study, these labels do not imply attached and detached flow, because upon closer inspection the surface pressure distributions show that neither state is fully attached or fully separated.
As a result, the $A$ is used to refer to the ``more attached'' state and the $D$ is used to refer to the ``more detached'' state.
The five free parameters ($\pi_A$, $c_{l,A}$, $c_{l,D}$, $\sigma_A$, $\sigma_D$) were estimated by maximum likelihood with the expectation-maximization algorithm, using the \texttt{fitgmdist} function in MATLAB.
The components were labeled by the magnitude of their means.
The samples were treated as independent draws, so the fit describes the distribution of the filtered record and not its dynamics.

Because a two-component mixture can be fit to any distribution, a separate criterion was used to decide whether a record is bimodal.
A record was classified as bimodal when the fitted mixture satisfied two conditions.
The first is that the two components are well separated relative to their widths, measured by Ashman's coefficient~\cite{Ashman1994},
\begin{equation}
D_{\mathrm{As}} = \frac{\left|c_{l,A} - c_{l,D}\right|}{\sqrt{\frac{\sigma_A^{2} + \sigma_D^{2}}{2}}} > 2,
\end{equation}
which compares the separation between the component means to their combined scatter and so excludes records in which a second component just fits the skewness or tail of a single state.
The second is that the fitted density has an antimode, that is, two distinct local maxima separated by a local minimum.
Because the first derivative of the mixture density is
\begin{equation}
\frac{d\mathrm{PDF}}{dc_l} = -\sum_{i \in \{A,D\}} \frac{\pi_i\,\left(c_l - c_{l,i}\right)}{\sigma_i^{3}\sqrt{2\pi}} \exp\!\left[-\frac{\left(c_l - c_{l,i}\right)^2}{2\sigma_i^2}\right],
\end{equation}
this condition requires a point $c_l^{*}$ between $c_{l,D}$ and $c_{l,A}$ at which $d\mathrm{PDF}/dc_l = 0$ and $d^2\mathrm{PDF}/dc_l^2 > 0$.
The first condition does not account for the weights of the components, and the second does not account for how well separated they are, so both are used.

Proper orthogonal decomposition (POD) was also applied to the unfiltered surface pressure fluctuations to identify the dominant spatial patterns of the unsteady loading, independent of the lift-based state classification.
For each combination of inflow, Reynolds number, angle of attack, and pitch direction, the fluctuating pressure coefficient at the 32 ports, with the temporal mean at each port removed, was decomposed as
\begin{equation}
C_p'(x_j,t) = \sum_{k=1}^{32} a_k(t)\,\phi_k(x_j),
\end{equation}
where $\phi_k$ are the spatial modes, ordered by the share of the variance each carries, and $a_k(t)$ are the corresponding time coefficients.
These values were computed from the singular value decomposition of the time-resolved pressure matrix with the MATLAB function \texttt{svd}.
No spatial weighting was applied, so the variance is that of the pressure coefficient summed over the ports.
The sign of each mode is arbitrary. 
In this study, $\phi_1$ was oriented so that it is negative over the suction side, so that positive $a_1$ amplifies the suction.
The time coefficient $a_1$ was classified into two states by the filtering, mixture model, and bimodality criterion applied to $c_l$ above.
The process for implementing and analyzing the POD is similar to that carried out by Nikoueeyan and Naughton~\cite{Nikoueeyan2023} and follows the insights and processes presented by Taira~\textit{et al.}~\cite{Taira2017, Taira2020}.

%============================================================
\section{Results and discussion}
\label{sec:results}
%%% Opening of Results, reiterate main idea and mention structure.
The aim of this study is to determine how freestream turbulence affects both the time-averaged and the unsteady loads on a thick, cambered airfoil near and beyond stall.
The results are presented in three parts.
Section~\ref{sec:meanlift} begins with the time-averaged lift and pressure drag presented for both the clean and turbulent inflows, the quantities by which airfoil performance is most commonly characterized.
Section~\ref{sec:bistability} resolves the lift coefficient in time to examine how freestream turbulence affects the unsteady loads that the time average conceals at and beyond the stall conditions.
Lastly, Section~\ref{sec:mechanism} examines the unsteady surface pressures to identify the physical origin of that unsteady loading and bistability.

%%%%%%%%%%%%%%%%%%%%%%%%%%%%%%%%%%%%%
%%% Mean (Time-Averaged) Response %%%
%%%%%%%%%%%%%%%%%%%%%%%%%%%%%%%%%%%%%
\subsection{Mean Aerodynamic Performance under Elevated Freestream Turbulence}
\label{sec:meanlift}
\begin{figure}[t]
  \centering
  \includegraphics[width=6.5in]{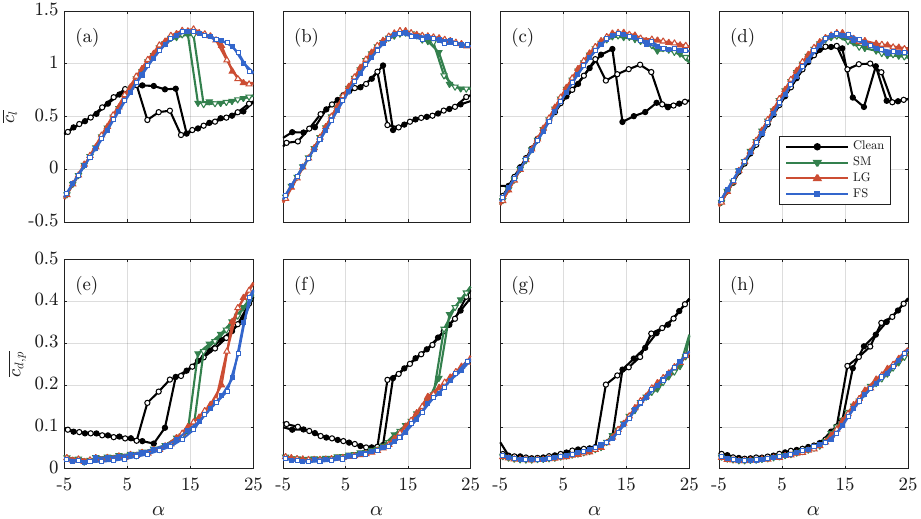}
  \caption{Time-averaged lift coefficient (a)--(d) and pressure drag coefficient (e)--(h) versus angle of attack for all configurations tested at $Re = 125{,}000$ (a, e), $175{,}000$ (b, f), $225{,}000$ (c, g), and $275{,}000$ (d, h).
 Filled markers denote the pitch-up portion of testing, and open markers the pitch-down portion.}
  \label{fig:polars}
\end{figure}

The instantaneous surface pressure was measured across the four inflow configurations of varying turbulence level and structure, and at the four chord-based Reynolds numbers between $Re = 125{,}000$ and $275{,}000$.
Data were collected over $\alpha = -30^\circ$ to $30^\circ$ in $0.9^\circ$ increments, through both pitch-up and pitch-down portions of testing to capture hysteresis in the pressure load response.
Figure~\ref{fig:polars} presents the resulting time-averaged lift coefficient $\overline{c_l}$ (a -- d) and pressure drag coefficient $\overline{c_{d,p}}$ (e -- h) versus angle of attack for the sixteen unique combinations.
The range $-5^\circ \leq \alpha \leq 25^\circ$ is shown, spanning the linear lift regime, stall, and the post-stall regime.
Note that the negative portion of the lift and drag polars (i.e., $-30^\circ \leq \alpha \leq -5^\circ$) is not shown here for brevity but shows findings consistent with those observable in Fig.~\ref{fig:polars} and discussed below.

In Fig.~\ref{fig:polars}, the clean configuration (black lines, i.e., no turbulence grid installed) is immediately distinguishable from the turbulent configurations.
Most notably, at $Re = 125{,}000$ and $175{,}000$, the lift slope of the clean airfoil from $-5^\circ \leq \alpha \leq 5^\circ$ is approximately 0.045 per degree, less than half the lift slope (0.10 per degree) measured for the turbulent cases.
This dramatically reduced lift slope at these lower Reynolds numbers is consistent with flow separation occurring on both the pressure and suction sides of the 30\%-thick FFA-W3-301 airfoil, which was observable in the pressure distributions but is not shown here for brevity.
Between $Re = 175{,}000$ and $225{,}000$, the character of the linear lift-slope regime changes, as the lift slope increases to 0.1 per degree and the pressure drag at $\alpha = 0^\circ$ falls from approximately 0.085 to 0.025, agreeing closely with the observed values for the elevated turbulence cases.
Note that this behavior remains consistent as the Reynolds number is further increased to $Re = 275{,}000$.
An abrupt change of this kind is consistent with a transition in boundary-layer state, from one unable to remain attached at low angles of attack to one that remains attached up to stall~\cite{Lissaman1983}.
With the increase in Reynolds number, the maximum lift coefficient of the clean configuration rises steadily to approximately $c_{l,\max} = 1.15$ at $Re = 275{,}000$ but remains below the turbulent range, which peaks at approximately $c_{l,\max} = 1.3$.
Interestingly, even at the lowest Reynolds numbers, the turbulent cases appear to reach a boundary-layer state comparable to the clean airfoil at much higher Reynolds numbers, as evidenced by the similar lift slopes and pressure drag between the different combinations of test conditions.

The cases of elevated turbulence display far more similarity to each other up to $c_{l,\max}$.
Through the linear regime, the polars of all three turbulence grids collapse onto a single curve at every Reynolds number.
The lift slopes are consistently approximately 0.1 per degree, and the pressure drag at $\alpha = 0^\circ$ is between 0.021 and 0.027, which is within the measurement uncertainty for these experiments.
The maximum lift coefficient is similarly insensitive to both the inflow and the Reynolds number.
It lies within $1.27 \leq c_{l,\max} \leq 1.33$ in every turbulent case, occurring within one angular increment of $\alpha = 14.6^\circ$.
At the lowest Reynolds number, the turbulent cases produce a linear-lift-regime response that the clean case reaches only at $Re \geq 225{,}000$, and a maximum lift that the clean case does not reach at any Reynolds number tested in the current study.
Up to maximum lift, the three grids are nearly identical, and the differences between them are confined to the post-stall regime.

Beyond the angle of attack at which the maximum lift coefficient is reached, the turbulent polars begin to show disagreement.
Some cases lose lift abruptly after $c_{l,\max}$ and move onto a low-lift, high-drag branch of the polar.
Other cases follow a gradual post-stall decline through the largest angle tested and remain on the low-drag branch of the polar.
The abrupt loss of lift is most pronounced for the SM grid at $Re = 125{,}000$, where $\overline{c_l}$ falls by 0.65, roughly half its maximum, between $c_{l,\max}$ and the next angle of attack, $\alpha = 16.2^\circ$.
As the Reynolds number increases, this loss of lift moves to progressively higher angles of attack and becomes less abrupt.
At $Re = 275{,}000$, no cases stall abruptly within the range of angles tested.

Increasing the turbulence intensity has a similar effect to increasing the Reynolds number. 
At $Re = 125{,}000$, the LG and FS grids show a similar loss-of-lift behavior at high angles of attack ($\alpha \approx 20$--$23^\circ$), spread over several angle-of-attack increments rather than concentrated within one increment.
This more gradual decline is consistent with trailing-edge stall, in which the separation point advances upstream from the trailing edge as the angle of attack increases~\cite{McCullough1951,Broeren1998}, and closely matches the stall behavior reported for the FFA-W3-301 at $Re = 1.6\times10^6$~\cite{Fuglsang1998}.
The abrupt loss of lift, observed for the SM grid at the lower turbulence intensity, is more consistent with leading-edge stall, in which the flow separates abruptly from a position very close to the leading edge and remains separated aft of this point.
Therefore, for the FFA-W3-301, freestream turbulence changes the character of stall more than its onset, for the conditions studied here, since all of the configurations were able to reach the same maximum lift coefficient condition.
This trend directly suggests that the primary role of elevated freestream turbulence is to achieve a lift and drag response for an airfoil that is representative of testing at a higher effective Reynolds number in a clean flow configuration.
This is believed to result from early bypass transition in the airfoil boundary layer imposed by the elevated freestream turbulence, which creates a boundary layer at a lower mean Reynolds number whose response more closely resembles that of a higher mean Reynolds number.
This is effectively similar to using a physical trip on the airfoil surface, but without the negative consequences of fixing a trip to a specific chordwise position and with a specific geometric scale, which can cause premature flow separation, among other unintended consequences~\cite{Fuglsang1998}. 

At $Re = 125{,}000$, the three turbulent cases in Fig.~\ref{fig:polars}(a) lose lift at different angles of attack, providing an indication of the relative roles of turbulence intensity and integral length scale.
The abrupt loss of lift occurs first for the SM grid, immediately after $c_{l,\max}$ at $\alpha = 16.2^\circ$, and later for the LG and FS grids at $\alpha = 19.9^\circ$ and $23.5^\circ$, respectively, each following a gradual decline.
Because the SM grid has the lowest intensity (2.5\% against approximately 4\% for the LG and FS grids; Table~\ref{tab:inflow}), this ordering is consistent with turbulence intensity being the primary driver in delaying the abrupt loss of lift.
The LG and FS grids share nearly the same turbulence intensity yet lose lift at different angles, indicating that integral length scale, or structure of the turbulence, also plays a role at this Reynolds number.
This distinction between the stall behavior of the LG and FS grids largely disappears at higher Reynolds numbers, although the LG grid retains slightly higher post-stall lift at $Re = 275{,}000$ (Fig.~\ref{fig:polars}(d)), a difference discussed in more detail in Sec.~\ref{sec:bistability} below.

Another distinction of the turbulent cases is the lack of hysteresis in the polars.
Figure~\ref{fig:polars} shows the polars from the pitch-up (filled markers) and pitch-down (open markers) portions of testing at each configuration and Reynolds number.
For the clean airfoil, lift collapses and recovers at different angles of attack in the two directions at every Reynolds number, producing a clear hysteresis loop.
Under freestream turbulence, however, no hysteresis is observed at any of the Reynolds numbers tested, with the pitch-up and pitch-down data collapsing onto a single curve.
For the SM grid at $Re = 125{,}000$, where the loss of lift is most abrupt, lift collapses between $\alpha = 14.6^\circ$ and $16.2^\circ$ during pitch-up and recovers as the angle decreases from $17.1^\circ$ to $15.5^\circ$ during pitch-down.
Therefore, there is likely a single transition angle between $15.5^\circ$ and $16.2^\circ$, and the hysteresis loop observed is narrower than the $0.9^\circ$ angular increment used in this study.
Hysteresis near stall is conventionally interpreted as the signature of two flow states that coexist over a range of angles of attack, with the flow remaining in whichever state it occupies as the angle changes~\cite{Mueller1985,Hristov2018}.
At the higher Reynolds numbers, the clean airfoil shows these two flow states of post-stall lift levels directly.
The absence of hysteresis under turbulence, however, does not by itself establish that only one state exists, a distinction that is largely concealed by the mean polar.
The same mean polar would result if both states were present and the flow alternated between them within each record, so that the pitch-up and pitch-down records sample the same mixture.
Atallah \textit{et al.} found this behavior for a NACA~0012 near stall, where intermittent switching between states removed the expected hysteresis loop~\cite{Atallah2024}.

\subsection{Emergence and Persistence of Bistable Switching in the Stall Regime}
\label{sec:bistability}
\begin{figure}[t]
  \centering
  \includegraphics[width=6.5in]{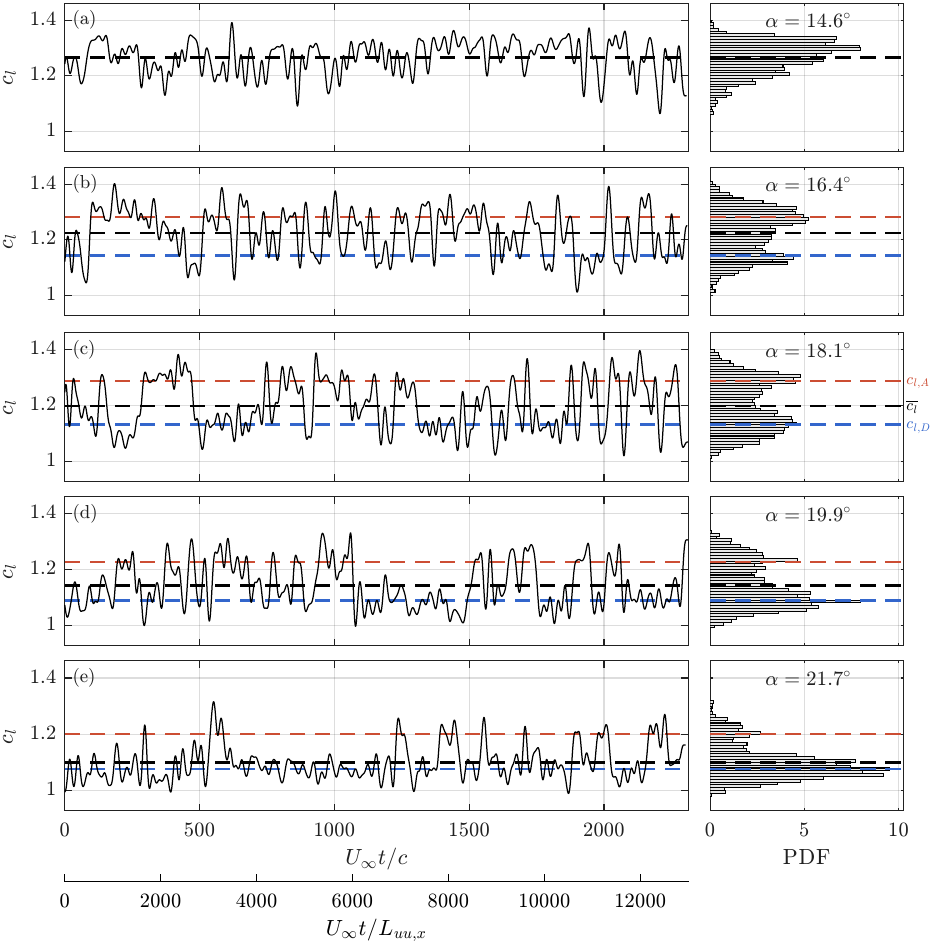}
   \caption{3~Hz low-pass filtered lift coefficient for the fractal square grid (FS) at $Re = 275{,}000$ during the pitch-up portion of testing.
   Rows (a)--(e) correspond to the angles of attack labeled in the right panels; left panels show the time series and right panels the corresponding probability density.
   Time is nondimensionalized by the chord on the upper axis, $U_\infty t/c$, and by the streamwise integral length scale of the inflow on the lower axis, $U_\infty t/L_{uu,x}$, with $L_{uu,x} = 45.9$~mm (Table~\ref{tab:inflow}).
   Dashed lines mark the higher-lift state $c_{l,A}$ (red), the lower-lift state $c_{l,D}$ (blue), and the time-averaged lift coefficient $\overline{c_l}$ (black); no state levels are shown in (a), where the distribution is unimodal.}
  \label{fig:timeseries}
\end{figure}

The time-averaged polars of Fig.~\ref{fig:polars} assign a single lift value to each angle of attack, but resolving the lift coefficient in time reveals dynamics that the time average conceals.
Figure~\ref{fig:timeseries} shows the 3~Hz low-pass filtered lift coefficient and its probability density for the fractal square grid at $Re = 275{,}000$ during the pitch-up portion of testing across five angles of attack spanning $\alpha = 14.6^\circ$ to $21.7^\circ$.
At $\alpha = 14.6^\circ$, the lift coefficient fluctuates about a single level and its distribution is unimodal, with a tail toward lower lift coefficient values.
At higher angles, the distribution is no longer centered on a single value but is split between two distinct lift levels, and the time series alternates between them.

The onset of the two-state behavior occurs near the stall angle of $\alpha = 14.6^\circ$.
By $\alpha = 16.4^\circ$, the distribution of the lift coefficient satisfies the bimodality criterion of Sec.~\ref{sec:analysis} and is well described by a two-component Gaussian mixture, with the record divided between a higher-lift state $c_{l,A}$ and a lower-lift state $c_{l,D}$.
Switching is therefore not confined to deep stall but appears soon after the airfoil departs from the linear lift regime.
At $\alpha = 16.4^\circ$ and $18.1^\circ$, the two states are separated by $\Delta c_l = 0.14$ and $0.16$, or 11\% and 13\% of the time-averaged lift coefficient, while $\overline{c_l}$ itself falls by 0.03 between the two angles.
The decline in $\overline{c_l}$ over this interval is therefore not driven by the states moving closer together or farther apart, but by a change in how the record is divided between them.

Beyond the onset of the bistable regime, the flow occupies the two states in increasingly unequal proportion.
The shift is most readily seen in the probability densities of Fig.~\ref{fig:timeseries}, where the weight of the distribution moves progressively from the upper mode to the lower one, and it appears in the time series as a growing fraction of the record spent near $c_{l,D}$.
At $\alpha = 16.4^\circ$, the two modes are comparable in weight, with the higher-lift state slightly favored; by $\alpha = 18.1^\circ$ the balance has reversed, and by $\alpha = 21.7^\circ$ the lower-lift $c_{l,D}$ mode dominates the distribution while excursions to $c_{l,A}$ have become brief and infrequent.
The same progression is reflected in the magnitude of $\overline{c_l}$, which is slightly less than $c_{l,A}$ at $\alpha = 16.4^\circ$ and is nearly equal to $c_{l,D}$ at $\alpha = 21.7^\circ$.
The state levels themselves are nearly unchanged over the first part of this range.
Over $16.4^\circ \leq \alpha \leq 18.1^\circ$, $c_{l,A}$ and $c_{l,D}$ each change by less than $0.02$, so the decline in $\overline{c_l}$ over that interval is almost entirely a redistribution of time spent within each of these fixed levels.
Beyond $\alpha = 18.1^\circ$, both state levels fall as well, and over the full range $16.4^\circ \leq \alpha \leq 21.7^\circ$, the two effects contribute comparably to the total decrease in $\overline{c_l}$.
The lift coefficient in the post-stall bistable regime is therefore set by two quantities that vary independently: (1) the lift levels of the two states, and (2) the fraction of time the flow spends within each state.

Because the load alternates between two levels rather than fluctuating about one, the mean of the time series gives a misleading description of the post-stall loading.
The time-averaged lift coefficient lies between the two states rather than being aligned with either of them.
Therefore, $\overline{c_l}$ is descriptive of a loading value that the system passes through rather than a state about which the lift is sustained.
This holds through most of the bistable range; only at $\alpha = 21.7^\circ$, where the record is dominated by the lower-lift state, does $\overline{c_l}$ fall within the scatter of a single state.

The standard deviation of the time series is similarly difficult to interpret, because it is set mainly by the distance between the two states rather than by the fluctuations observed within either state.
Over the bistable range, the separation between the states contributes roughly 60--70\% of the variance of the filtered lift coefficient, while the scatter within the states changes little with angle of attack.
The unsteady loading in this regime is therefore described more accurately by the two state levels and the fraction of time spent in each rather than by conventional statistics of the whole record, such as the mean and standard deviation of the full time series.

\begin{figure}[t]
  \centering
  \includegraphics[width=6.5in]{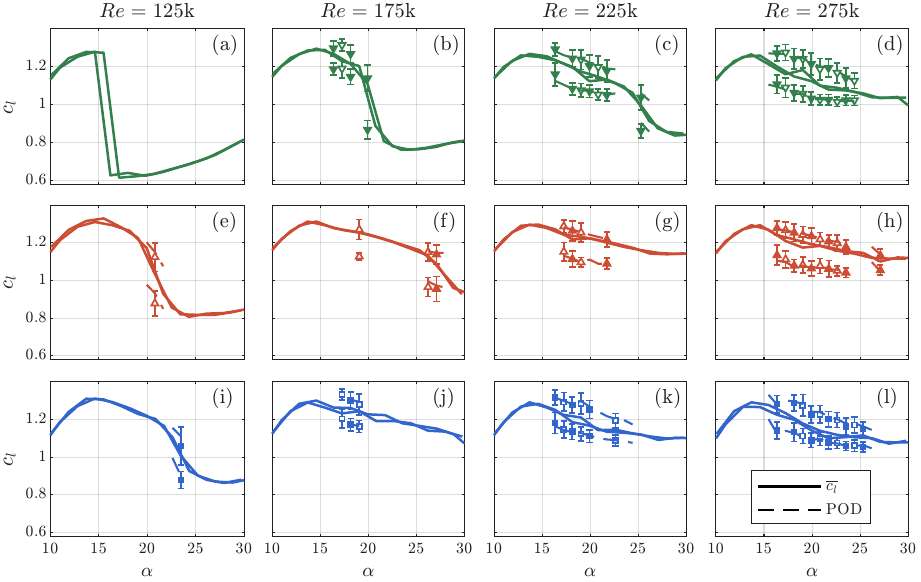}
  \caption{Time-averaged lift coefficient $\overline{c_l}$ (solid lines) versus angle of attack for the SM (a--d, $\triangledown$), LG (e--h, $\triangle$), and FS (i--l, $\square$) grids at $Re = 125{,}000$, $175{,}000$, $225{,}000$, and $275{,}000$ (left to right).
  At angles of attack where the 3~Hz low-pass filtered $c_l$ satisfies the bimodality criterion, markers show the higher-lift state ($c_{l,A}$, upper branch) and lower-lift state ($c_{l,D}$, lower branch) lift coefficients, with bars marking one standard deviation of each mixture component.
  Filled markers denote the pitch-up portion of testing and open markers the pitch-down portion; the solid $\overline{c_l}$ lines are likewise drawn separately for each pitch direction.
  Dashed lines are the state means obtained by conditioning $c_l$ on the POD mode~1 time coefficient $a_1$, shown where $a_1$ satisfies the same bimodality criterion, with both pitch directions shown as a single curve per state.}
  \label{fig:statemeans}
\end{figure}

Figure~\ref{fig:timeseries} establishes the two-state behavior for a single configuration at a single Reynolds number and at five angles of attack in a single pitch direction.
To determine whether it is a general feature of the stalled airfoil under elevated turbulence rather than a property of one case, the state identification of Sec.~\ref{sec:analysis} is applied to all of the turbulent freestream cases for every grid configuration, Reynolds number, angle of attack, and pitch direction in the dataset.
The results of this analysis are shown in Fig.~\ref{fig:statemeans}, where solid lines are the time-averaged lift coefficient, drawn separately for the pitch-up and pitch-down portions of testing so that any hysteresis remains visible.
At angles of attack where the lift coefficient satisfies the bimodality criterion, markers give the two state levels $c_{l,A}$ and $c_{l,D}$ obtained from the two-component Gaussian mixture model, with error bars spanning one standard deviation for each specific state.
Angles at which the two-state criterion is not met (refer to Sec.~\ref{sec:analysis}) carry no markers.
Figure~\ref{fig:statemeans} shows that bimodality is not confined to only the cases shown in Fig.~\ref{fig:timeseries} and instead is present for every turbulent configuration at $Re \geq 175{,}000$ and in both pitch directions, denoted by filled markers (pitch-up) and open markers (pitch-down), respectively.

An apparent feature of Fig.~\ref{fig:statemeans} is the growth in the extent of the bistable regime as a function of the Reynolds number.
At $Re = 275{,}000$, the bimodality criterion is satisfied over a range of roughly $5^\circ \leq \Delta\alpha \leq 9^\circ$ for every turbulence configuration, beginning near $\alpha = 16^\circ$ in the pitch-up portion of testing and persisting past $\alpha = 23^\circ$.
Switching is therefore not a narrow feature localized at the stall angle but a regime that occupies a large region of the post-stall polar.
This region is significantly broader than what has been reported for thin airfoils. 
Specifically, Atallah \textit{et al.}~\cite{Atallah2024} identify a bistable range of $\Delta\alpha \approx 0.2^\circ$ on a NACA~0012 in clean flow, more than an order of magnitude narrower than the range observed here for the FFA-W3-301 with elevated freestream turbulence.

Although the extent of the bistable regime is similar across the three grids, the time spent in each state is not.
At $Re = 275{,}000$ and at the angles of attack common to all three configurations, the state levels themselves are nearly independent of the inflow.
The two states, $c_{l,A}$ and $c_{l,D}$, agree across the grids to within roughly one standard deviation of the mixture components, and the separation between them follows the same variation with angle of attack in every case.
The time spent in each state, however, is not consistent.
At $\alpha = 21.7^\circ$, the flow spends 79\% of the time in the higher-lift state for the LG grid, 32\% for SM, and 19\% for FS, so the time-averaged lift coefficient lies near $c_{l,A}$ in the LG case and near $c_{l,D}$ in the other cases while the two state levels are nearly identical.
The progression of the states with angle of attack also differs.
For the SM and FS grids, the fraction of time spent in the higher-lift state falls steadily through the bistable regime, as in Fig.~\ref{fig:timeseries}, whereas for the LG grid it remains between 74\% and 84\% across the entire band.
The decline of $\overline{c_l}$ with angle of attack is therefore produced mainly by a redistribution of time between the states for the SM and FS cases, but almost entirely by the drift of the state levels themselves for the LG cases.
To understand whether this trend is a unique property of the turbulence structure produced by the LG grid or a unique result of this specific test run would require more experimental repetitions, which were deemed outside the scope of the current study given the consistency uncovered in the bistable state identification across all of the turbulence grids.

It is worth noting that the behavior described above is resolved only at the higher Reynolds numbers tested in this study.
At $Re = 125{,}000$, no bands of bimodal lift are found for any of the grid configurations.
More specifically, the SM grid (lowest turbulence intensity) produces no angles that satisfy the bimodality criterion at all, while the LG and FS grids (higher turbulence intensity) each produce a single isolated angle at which bimodality is observed.
At $Re = 175{,}000$ and $225{,}000$, the bimodality criterion is met at more angles, but the passing angles are scattered and the contiguous bands that do form only span $1.8^\circ \leq \Delta\alpha \leq 5.4^\circ$, at most two-thirds the width found at the highest Reynolds number.
Only at $Re = 275{,}000$ does every configuration begin to produce an unbroken band of bimodal lift signatures.
The width of the contiguous band therefore appears to grow monotonically with Reynolds number in each grid.

The results of this section establish that the stalled FFA-W3-301 under elevated freestream turbulence does not settle into a single state but alternates between two distinct lift levels over a broad band of post-stall angles of attack.
At $Re \geq 225{,}000$, the lift coefficients of the two states are nearly identical across the three turbulence grids evaluated, so the state levels appear insensitive to the inflow over the range of turbulence intensities and integral length scales tested.
However, the fraction of time spent within each state is less uniform and shifts toward the lower-lift state with increasing angle of attack and also differs between the grids at any given angle of attack.
While not shown here, the clean configuration (i.e., no freestream turbulence) behaved quite differently.
Applying the same criterion to the clean configuration across all Reynolds numbers returned no more than two isolated angles at which a bimodal lift state was observed, for any given Reynolds number, and no contiguous band at any of them.
This result does not establish that switching is absent in clean inflow.
A record containing a single transition can satisfy the criterion without demonstrating repeated switching, and a record containing none cannot satisfy it at all.
Without elevated freestream turbulence to supply the disturbances that trigger transitions, switching may be too infrequent for a 30~s record to capture it.
The present measurements show that, under elevated freestream turbulence, the flow visits both states repeatedly within a single 30~s record, approximately $2{,}300$ convective times at $Re = 275{,}000$.
In the low-disturbance experiments carried out by Atallah \textit{et al.}, by contrast, residence times alone were of order $10^4$ convective times, and records of 30~min or longer were required to build sufficient state-switching statistics~\cite{Atallah2024}.
Switching is therefore far more frequent than previously reported, yet each residence in a state still spans many passage times of the inflow integral scale, and therefore the transition observed in this study still remains rare relative to the advection of the eddies that force it.
This is consistent with the rare-event description of Atallah \textit{et al.}~\cite{Atallah2024}, but with freestream turbulence supplying the triggering disturbances far more often than a low-disturbance inflow does.

\subsection{Physical Source of Bimodality}
\label{sec:mechanism}
\begin{figure}[t]
  \centering
  \includegraphics[width=6.5in]{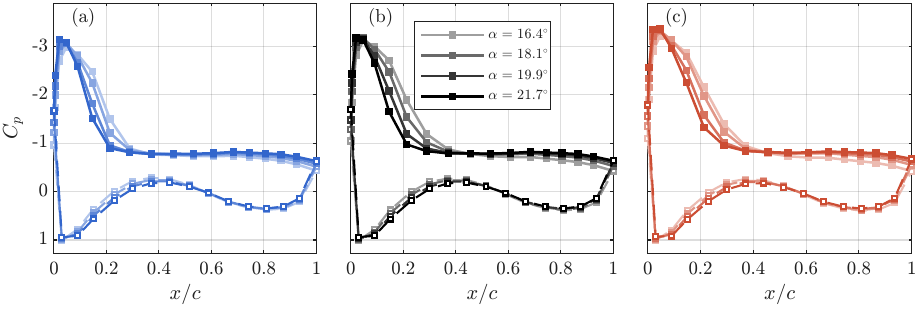}
  \caption{State-conditioned mean pressure coefficient distributions for the FS grid at $Re = 275{,}000$ during the pitch-up portion of testing: (a) lower-lift state, (b) unconditioned time average, and (c) higher-lift state.
  Solid lines with filled squares denote the suction side; dashed lines with open squares denote the pressure side.
  Line shades darken with increasing angle of attack.}
  \label{fig:cpstates}
\end{figure}

Section~\ref{sec:bistability} established that the flow alternates between two lift levels, but treated those levels only as values extracted from the lift coefficient PDF.
What physically distinguishes the two states is examined here through a deeper analysis of the surface pressure distributions.
Assigning each sample of the record to a state using the 3~Hz low-pass filtered lift coefficient, and then conditionally averaging the raw surface pressures within one standard deviation of the state's Gaussian peak, separates the time-averaged pressure coefficient distribution into the two distributions the flow actually occupies.
Unlike the lift coefficient, which is a single integrated quantity, the conditioned distributions resolve where along the chord the two states differ.
Figure~\ref{fig:cpstates} shows the result for the FS grid at $Re = 275{,}000$ during the pitch-up portion of testing.
Results are shown at the same four angles of attack that satisfied the bimodality criterion presented in Fig.~\ref{fig:timeseries} and are presented alongside the unconditioned time average (Fig.~\ref{fig:cpstates}b) for reference.

\begin{figure}[t]
  \centering
  \includegraphics[width=6.5in]{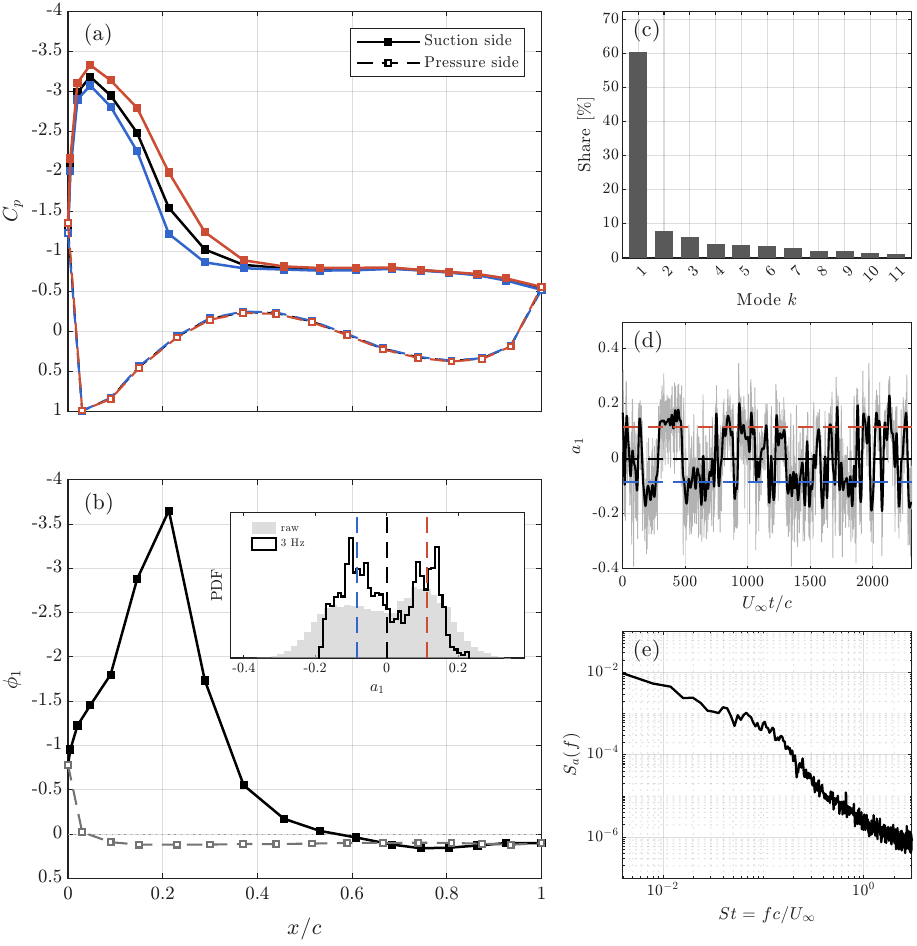}
  \caption{Proper orthogonal decomposition of the unfiltered surface pressure fluctuations for the FS grid at $Re = 275{,}000$ and $\alpha = 18.1^\circ$ during the pitch-up portion of testing.
  (a) State-conditioned mean pressure coefficient distributions: lower-lift state (blue), unconditioned time average (black), and higher-lift state (red).
  (b) First spatial mode $\phi_1$; the inset shows the probability density of its time coefficient $a_1$, raw (shaded) and 3~Hz low-pass filtered (outlined).
  (c) Share of the total pressure variance carried by each mode, shown up to 95\% cumulative variance.
  (d) Time series of $a_1$, raw (gray) and 3~Hz low-pass filtered (black).
  (e) Power spectral density of $a_1$ versus chord-based Strouhal number.
  In (a) and (b), solid lines with filled squares denote the suction side and dashed lines with open squares the pressure side.
  Dashed lines in the inset of (b) and in (d) mark the unconditioned and state-conditioned means of $a_1$, colored as in (a).}
  \label{fig:pod}
\end{figure}

The two states display a similar pressure topology at every angle of attack in the bistable range, as shown in Fig.~\ref{fig:cpstates}.
In both states and the time-averaged data, a suction peak forms at the leading edge, the pressure recovers over the forward portion of the suction side, and the recovery ends on a nearly constant-pressure plateau that extends from $x/c \approx 0.4$ to the trailing edge.
The plateau levels of the two states agree to within $\Delta C_p = \pm0.1$ at every angle of attack from $x/c \approx 0.4$ to the trailing edge, where $C_p$ changes from $C_p \approx -0.7$ at $\alpha = 16.4^\circ$ to $C_p \approx -0.8$ at $\alpha = 21.7^\circ$. 
As a result, the flow is separated over the aft chord in both states.
In both states, the flow consistently remains attached over the forwardmost chordwise positions, $x/c \leq 0.1$. 
What differs between the two states is how far the pressure recovery extends before the plateau is reached and the pressure coefficient magnitude achieved at the suction peak.
The plateau begins at $x/c = 0.53$ in the higher-lift state (Fig.~\ref{fig:cpstates}c) compared with $x/c = 0.37$ for the lower-lift state at $\alpha = 16.4^\circ$, and at $x/c = 0.37$ compared with $x/c = 0.29$ at $\alpha = 21.7^\circ$.
The difference between the two distributions is largest in the pressure recovery region, which is most distinctly observed by comparing the values of $C_p$ between the lower- and higher-lift states (Fig.~\ref{fig:cpstates}a \& c, respectively) over $0.1 \leq x/c \leq 0.4$.
The higher-lift state also sustains a stronger suction peak, with a minimum pressure coefficient of $C_{p,\min} = -3.22$ compared with $C_{p,\min} = -2.98$ at $\alpha = 16.4^\circ$ and $C_{p,\min} = -3.37$ compared with $C_{p,\min} = -3.14$ at $\alpha = 21.7^\circ$.
That said, the difference in magnitude of the suction peak is less than 10\% of the local pressure coefficient at that location, whereas across the recovery region the two distributions differ by roughly half the local pressure coefficient value, so the primary difference between the states can be classified by the extent of the pressure recovery region just downstream of the suction peak.
It should also be noted that the pressure side of the airfoil is nearly identical between the two states; as a result, the difference between the states arises almost entirely on the suction side.
The change between higher-lift and lower-lift states is therefore not the product of a dramatic change in the flow topology (i.e., transitioning from fully attached to fully separated flow) but rather the more subtle motion or intermediate location of the separation point along the suction surface.

Interestingly, within either state (lower- or higher-lift), the airfoil responds to increasing angle of attack consistent with the conventional understanding of trailing-edge stall.
The plateau onsets given above each move upstream by one to two pressure ports over $16.4^\circ \leq \alpha \leq 21.7^\circ$, so the separation point advances with angle of attack even for a single flow state.
The suction peak strengthens over the same range, partly offsetting the loss of lift from earlier separation, so the state levels remain nearly constant up to $\alpha = 18.1^\circ$ and fall beyond it.
The decrease of $\overline{c_l}$ through the bistable regime therefore combines two distinct behaviors.
One is the advance of separation within each state, which sets the state levels, and the other is the redistribution of time spent by the flow within each of the states, as described in Sec.~\ref{sec:bistability}.
The pressure distributions are shown for the FS grid only, but the state levels and the state pressure distributions agree across all three grids (Sec.~\ref{sec:bistability}), which indicates that the behavior within each state is common and that the grids differ mainly in how much time is spent in the states.
More specifically, for the LG grid, which remains predominantly in the higher-lift state across the bistable range, the advance of flow separation within that higher-lift state accounts for nearly all of the decrease in $\overline{c_l}$.

The states of Fig.~\ref{fig:cpstates} were identified from the filtered lift coefficient, so it remains to be shown that the pressure field divides into the same two states when no such classification is imposed.
If the two states are two positions of a single separation front, as the plateau onsets of Fig.~\ref{fig:cpstates} suggest, the fluctuating surface pressures should be dominated by one spatial pattern whose amplitude alternates between these two levels.
Proper orthogonal decomposition (POD) provides a direct test of this expectation by decomposing the raw pressure fluctuations into spatial patterns ranked by the share of the variance each carries.
The POD makes no reference to the states, so although the lift coefficient is integrated from the same pressures, any two-state structure it recovers is independent of the thresholds and filters used to obtain the state classification.
Figure~\ref{fig:pod} shows the result for the FS grid at $Re = 275{,}000$ and $\alpha = 18.1^\circ$ during the pitch-up portion of testing.
The state-conditioned distributions at this angle are repeated in Fig.~\ref{fig:pod}a for comparison with the first spatial mode $\phi_1$ in Fig.~\ref{fig:pod}b, while Figs.~\ref{fig:pod}c and \ref{fig:pod}d show how the variance is distributed across the modes and how the time coefficient of the first mode, $a_1$, evolves over the time record.
Lastly, the spectrum of the time coefficient, $a_1$, is shown in Fig.~\ref{fig:pod}e.

POD largely confirms what is observed through the analysis of the lift coefficient.
The lift coefficient alternates between two states, and the pressure field switches through a single spatial pattern concentrated in the pressure recovery region.
Figure~\ref{fig:pod}c shows that the first mode accounts for 61\% of the pressure variance, nearly eight times the share of the second mode, while the remaining variance is spread across many modes.
Figure~\ref{fig:pod}b shows that the spatial shape of this mode is concentrated on the suction side, confirming what was observed in Fig.~\ref{fig:cpstates}.
Along the suction surface, the mode grows from the leading edge to a maximum at $x/c = 0.21$ and decays to near zero over most of the separated plateau, while on the pressure side it is nearly zero aft of the stagnation point, $x/c \geq 0.04$.
The maximum lies at the location of the largest difference between the state-conditioned pressure distributions in Fig.~\ref{fig:pod}a, and the spatial mode magnitude is weaker at the suction peak and over the separation plateau region, where the two states differ less.
As a result, $\phi_1$ follows the same distribution along the chord as the difference between the pressure distribution states observed in Fig.~\ref{fig:cpstates} and discussed above.
The time coefficient $a_1$ shows the corresponding two-state behavior in time, as shown in Fig.~\ref{fig:pod}d.
Its probability density is bimodal in the raw coefficient and more strongly so after 3~Hz low-pass filtering, with the antimode near the unconditioned mean (see inset plot in Fig.~\ref{fig:pod}b).
The two states identified from the lift coefficient therefore correspond to two amplitudes of a single pressure pattern, consistent with the end of the pressure recovery region, and with it the separation point, moving forward or aft along the suction surface as the flow changes states.

Figure~\ref{fig:pod} identifies the dominant pressure mode for a single grid, Reynolds number, and angle of attack.
The dashed lines in Fig.~\ref{fig:statemeans} extend the comparison to every configuration, giving the state levels obtained when each record is classified on $a_1$ instead of $c_l$ with the same bimodality criterion applied to both.
At the highest Reynolds numbers, the two state-classification methods are nearly interchangeable.
%At $Re \geq 225{,}000$, in the two right-hand columns of Fig.~\ref{fig:statemeans} [panels (c), (d), (g), (h), (k), and (l)], they assign a 96\% of samples to the same state, and the resulting state levels differ by a median of 3\% of the separation between the states.
More specifically, the two classifications differ mainly in how many angles they identify as bimodal.
Every angle that satisfies the bimodality criterion in $c_l$ also satisfies it in $a_1$, whereas $a_1$ satisfies the bimodality criterion for an additional 14 angles of attack, 12 of which extend or join the existing bands.
The lift coefficient is the less sensitive variable because it integrates the contribution from every mode of the pressure field; thus, the POD appears to be more sensitive to directly isolating the state switching phenomena.

What carries the flow from one state to the other is examined through the time behavior of $a_1$ for the fractal square grid at $Re = 275{,}000$.
The spectrum of $a_1$ presented in Fig.~\ref{fig:pod}e for $\alpha = 18.1^\circ$ contains no dominant frequency peak, nor was any observed for any of the other bimodal angles presented in Fig.~\ref{fig:timeseries}.
As a result, the switching is not the signature of an oscillation with a defined frequency, such as the low-frequency stall oscillation.
Instead, this memoryless switching resembles the behavior reported by Atallah \textit{et al.} for a NACA~0012 at stall for a clean inflow, which they describe as a Markov process whose transitions are triggered by rare, extreme fluctuations attributed to the background turbulence within their facility~\cite{Atallah2024,Boujo2025}.
Because the pressures are measured at a single spanwise station, the two states identified here are sectional states, and simultaneous measurements at several spanwise stations would be needed to further determine whether a separation front moves uniformly across the span or within spanwise-localized structures such as stall cells.

% ============================================================
\section{Conclusions}
\label{sec:conclusions}
The aerodynamics of a thick, cambered airfoil section, near and beyond stall, in a turbulent inflow was examined through time-resolved surface pressure measurements at the midspan of an FFA-W3-301 airfoil model to address three primary research questions as part of this study. 
First, the influence of freestream turbulence on the mean lift and pressure drag was assessed across a range of Reynolds numbers between $125{,}000$ and $275{,}000$, to ascertain whether the response is governed solely by turbulence intensity or also by integral length scale. 
Second, the distributions of the lift over time were analyzed in detail for a range of angles of attack near and beyond stall to quantify the range of angles over which the lift switches between two distinct states, and to determine whether turbulence (intensity and structure) alters the lift levels of the states or the residence time in each state.
Finally, the spatial variations in the pressure distribution were conditionally sampled and decomposed to determine how the pressure distribution along the chord differs between the two states.
This decomposition was used to isolate whether the separation point jumps between distinct positions or advances progressively along the chord.

From the mean lift and pressure drag analysis, it was found that the presence of freestream turbulence more significantly altered the mean response as compared with the choice of any particular turbulence grid itself.
Specifically, the clean airfoil was strongly Reynolds-number dependent.
At $Re = 125{,}000$ and $175{,}000$, its lift slope was less than half that for all of the turbulent freestream cases, and its pressure drag at $\alpha = 0^\circ$ was elevated, reflecting flow separation on both the pressure and suction surfaces well below stall.
The clean airfoil reached a comparable linear regime only at $Re \geq 225{,}000$, and its maximum lift coefficient remained notably below the turbulent freestream values at every Reynolds number tested.
Under turbulent inflow, the mean lift and drag polars of all three grids collapsed through the linear regime for every Reynolds number tested, and $c_{l,\max}$ fell within $\pm 0.9^\circ$ of $\alpha_s = 14.6^\circ$ for every grid.
Among the turbulent freestream cases, the inflow altered the character of the post-stall response more than the onset of stall (i.e., $\alpha_s$ and $c_{l,\max}$).
At the lowest Reynolds number and turbulence intensity, lift was lost abruptly after $c_{l,\max}$, consistent with separation originating near the leading edge.
As either the Reynolds number or the turbulence intensity increased, this loss moved to higher angles of attack and became a gradual decline consistent with the trailing-edge stall reported for this section at $Re = 1.6 \times 10^6$~\cite{Fuglsang1998}.
These trends are consistent with turbulence-promoted transition producing a boundary layer that is representative of one at a higher effective Reynolds number, similar to the result expected for a boundary-layer trip, but without fixing the transition location.
Turbulence intensity appears to be the primary parameter governing this response, with the integral length scale, or the structure of the turbulence more broadly, playing a secondary role evident only at the lowest Reynolds number when comparing the differences between the fractal square and large classical grid cases, which had equivalent turbulence intensity.
Turbulence also removed the hysteresis exhibited by the clean airfoil for every Reynolds number tested, collapsing the pitch-up and pitch-down polars onto a single curve to within the resolution of the angular increment of $0.9^\circ$.

Resolving the lift coefficient in time showed that the gradual post-stall decline in the mean lift coefficient concealed a bistable switching between flow attachment states.
Beyond $c_{l,\max}$, the lift alternated repeatedly between a higher-lift and a lower-lift state.
The range of bistability was observed to grow with Reynolds number, reaching a maximum of $5^\circ \leq \Delta \alpha \leq 9^\circ$ for each of the turbulence grids (inclusive of both pitch directions) at the highest Reynolds number that was tested, namely $Re = 275{,}000$.
This range was found to be more than an order of magnitude wider than the bistable range of approximately $\Delta \alpha = 0.2^\circ$ reported for a thin, symmetric section in low-turbulence inflow~\cite{Atallah2024}.
That said, it is comparable to the range observed on a 20\%-thick, cambered section in low-turbulence inflow at high Reynolds numbers~\cite{Neunaber2022}, suggesting that section geometry contributes to the width of this bistable switching range in angle of attack.
For the turbulence conditions tested, the lift levels of the two states were nearly independent of the grid at $Re \geq 225{,}000$, while the fraction of time spent in each state was found to differ between grids.
The decline of the time-averaged lift coefficient through the bistable regime combines a drift of the state levels common to every grid with a redistribution of time spent in the states, which depended on the turbulence conditions.
Because the lift alternates between two levels, the mean and standard deviation of the record remain valid statistics but cannot be read in the conventional sense of a typical load and a fluctuation amplitude about it.
More specifically, the mean was found to lie between the two states, at a value the flow passes through rather than one the flow occupies, and the standard deviation is set largely by the separation between the states, which accounts for 60--70\% of the variance of the filtered lift coefficient, rather than by the fluctuations within either state.
Repeated switching also explains the absence of hysteresis under turbulent inflow: the lift changed state many times within each recorded time period, so the pitch-up and pitch-down portions of testing sufficiently sampled the same distribution to demonstrate a consistent trend in the mean response.

To better understand the physical response of the flow field, the conditionally averaged surface pressure distributions were used to distinguish the two distinct lift states.
Throughout the angle-of-attack range in which the bistable switching was observed, both states were found to be partially separated; neither was fully attached or fully separated (detached).
In fact, both states retained the leading-edge suction peak and ended in a separated pressure plateau over the aft chord region, with nearly identical pressure-side distributions; they differed primarily in how far pressure recovery on the suction side persisted before the plateau began, which was found to reach farther downstream in the higher-lift state.
Furthermore, through proper orthogonal decomposition, it was found that a single mode carried 61\% of the pressure variance, which was concentrated in the pressure recovery region on the suction side, and on its own was able to capture the switching.
Not surprisingly, the time coefficient for this mode was found to be bimodal, with a broadband spectrum and no characteristic switching frequency.
The reconstruction of the lift coefficient also matched (within the experimental uncertainty) the state levels found independently through the two-component Gaussian mixture model analysis across the range of bistable angles of attack.
These results suggest that the two states represent two positions of a single separation front along the suction surface of the airfoil.
At a fixed angle of attack, the separation point jumps between these locations, producing the bistable response observed. 
However, the separation position does retreat progressively upstream for each of the states with an incremental increase in angle of attack.
This picture is consistent with the intermittent separation-point jumps observed on a thick section for a low-turbulence inflow~\cite{Neunaber2022}, but it also suggests that the fluctuating separation point reported on thick airfoil sections in turbulent inflow~\cite{Sicot2006} moves between distinct positions rather than continuously along the chord, at least for the turbulence conditions tested within this study.

These results are relevant directly to wind turbine blades, whose sections are thick relative to those on which bistable stall has typically been studied and also operate continuously in turbulent inflow~\cite{Veers2019}. 
Blade element momentum models, the standard tool for predicting blade loads, typically assign a single lift coefficient to each angle of attack from a time-averaged polar. 
When blade sections operate near or beyond stall, such models cannot represent the switching between two lift states documented here and may therefore under-predict the unsteady loading and fatigue the blade experiences. 
Because the surface pressures were only measured at midspan of the airfoil model, the states identified within this study are only sectional, but the findings are consistent with integrated load measurements made by other researchers~\cite{Atallah2024}.
Simultaneous measurements at several spanwise stations would establish whether the separation front moves uniformly across the span or within spanwise local structures such as stall cells~\cite{Braud2024, Hanna2026}.
Extending these measurements to higher Reynolds numbers and a wider range of turbulence intensities would also help to determine if the bistable regime continues to broaden or change.

% ============================================================
\begin{acknowledgments}
This work is supported by the National Science Foundation (NSF) Graduate Research Fellowship Program (Grant No.: 2040434).
\end{acknowledgments}

%\appendix

\bibliography{refs}

@PREAMBLE{
 "\providecommand{\noopsort}[1]{}" 
 # "\providecommand{\singleletter}[1]{#1}%" 
}

@article{Atallah2024,
  title={From low-frequency oscillations to Markovian bistable stall dynamics},
  author={Kharsansky Atallah, Ivan and Pastur, Luc and Monchaux, Romain and Zimmer, Laurent},
  journal={Physical Review Fluids},
  volume={9},
  number={6},
  pages={063902},
  year={2024},
  publisher={APS},
  doi={10.1103/PhysRevFluids.9.063902}
  
}

@article{Boujo2025,
  title={Bistable flow dynamics of airfoil stall under varying angle of attack: A stochastic model with multiplicative noise},
  author={Boujo, Edouard and Kharsansky Atallah, Ivan and Pastur, Luc R},
  journal={Physical Review E},
  volume={112},
  number={3},
  pages={L032202},
  year={2025},
  publisher={APS},
  doi={10.1103/8jz3-3j3t}
}

@article{Lissaman1983,
  title={Low-Reynolds-number airfoils},
  author={Lissaman, PBS},
  journal={Annual review of fluid mechanics},
  volume={15},
  number={1},
  pages={223--239},
  year={1983},
  doi={10.1146/annurev.fl.15.010183.001255}
}

@techreport{McCullough1951,
  title       = {Examples of three representative types of airfoil-section stall at low speed},
  author      = {McCullough, George B. and Gault, Donald E.},
  institution = {National Advisory Committee for Aeronautics},
  type        = {Technical Note},
  number      = {NACA-TN-2502},
  address     = {Washington, DC},
  month       = sep,
  year        = {1951}
}

@inproceedings{Broeren1998,
  title={Low-frequency flowfield unsteadiness during airfoil stall and the influence of stall type},
  author={Broeren, Andy and Bragg, Michael},
  booktitle={16th AIAA Applied Aerodynamics Conference},
  pages={2517},
  year={1998},
  doi = {10.2514/6.1998-2517}
}

@techreport{Fuglsang1998,
  title       = {Wind Tunnel Tests of the {FFA-W3-241}, {FFA-W3-301} and {NACA} 63-430 Airfoils},
  author      = {Fuglsang, Peter and Antoniou, Ioannis and Dahl, Kristian S. and Madsen, Helge Aagaard},
  institution = {Ris{\o} National Laboratory},
  type        = {Ris{\o}-R Report},
  number      = {Ris{\o}-R-1041(EN)},
  address     = {Roskilde, Denmark},
  month       = nov,
  year        = {1998},
  isbn        = {87-550-2443-4},
}

@article{Farnsworth2020,
  title={Design and qualification of an unsteady low-speed wind tunnel with an upstream louver system},
  author={Farnsworth, J and Sinner, D and Gloutak, D and Droste, L and Bateman, D},
  journal={Experiments in fluids},
  volume={61},
  number={8},
  pages={181},
  year={2020},
  publisher={Springer},
  doi = {10.1007/s00348-020-03018-1}
}

@techreport{Gaertner2020,
  title       = {Definition of the {IEA} 15-Megawatt Offshore Reference Wind Turbine},
  author      = {Gaertner, Evan and Rinker, Jennifer and Sethuraman, Latha and Zahle, Frederik and Anderson, Benjamin and Barter, Garrett and Abbas, Nikhar and Meng, Fanzhong and Bortolotti, Pietro and Skrzypi{\'n}ski, Witold and Scott, George and Feil, Roland and Bredmose, Henrik and Dykes, Katherine and Shields, Matt and Allen, Christopher and Viselli, Anthony},
  institution = {National Renewable Energy Laboratory},
  address     = {Golden, CO},
  number      = {NREL/TP-5000-75698},
  year        = {2020},
  month       = mar,
  doi = {10.2172/1603478}
}

@techreport{Zahle2024,
  title       = {Definition of the {IEA} Wind 22-Megawatt Offshore Reference Wind Turbine},
  author      = {Zahle, Frederik and Barlas, Athanasios and L{\o}nb{\ae}k, Kenneth and Bortolotti, Pietro and Zalkind, Daniel and Wang, Lu and Labuschagne, Casper and Sethuraman, Latha and Barter, Garrett},
  institution = {Technical University of Denmark},
  address     = {Roskilde, Denmark},
  type        = {DTU Wind Report},
  number      = {E-0243},
  year        = {2024},
  doi={10.11581/DTU.00000317}
}

@article{BandT1975,
  title={On the propagation of sound waves in cylindrical tubes},
  author={Tijdeman, H},
  journal={Journal of Sound and Vibration},
  volume={39},
  number={1},
  pages={1--33},
  year={1975},
  publisher={Elsevier},
  doi = {10.1016/S0022-460X(75)80206-9}
}

@book{Barlow1999,
  title={Low-speed Wind Tunnel Testing},
  author={Barlow, Jewel B and Rae, William H and Pope, Alan},
  year={1999},
  publisher={John Wiley \& Sons}
}

@article{Ashman1994,
  title={Detecting bimodality in astronomical datasets},
  author={Ashman, Keith M and Bird, Christina M and Zepf, Steven E},
  journal={arXiv Preprint},
  year={1994}, 
  doi={10.48550/arXiv.astro-ph/9408030}
}

@article{Cao2011,
  title={The Performance of a High-Lift Airfoil in Turbulent Wind},
  author={Cao, N and Ting, D SK and Carriveau, R},
  journal={Wind Engineering},
  volume={35},
  number={2},
  pages={179--196},
  year={2011},
  publisher={SAGE Publications Sage UK: London, England}, 
  doi={10.1260/0309-524X.35.2.179}
}

@article{Saric2002,
  title={Boundary-layer receptivity to freestream disturbances},
  author={Saric, William S and Reed, Helen L and Kerschen, Edward J},
  journal={Annual review of Fluid Mechanics},
  volume={34},
  number={1},
  pages={291--319},
  year={2002},
  publisher={Annual Reviews},
  doi={10.1146/annurev.fluid.34.082701.161921}
}

@inproceedings{Yap2001,
  title={The effect of turbulence intensity on the aerodynamic performance of airfoils},
  author={Yap, TC and Abdullah, MZ and Husain, Z and Ripin, Z Mohd and Ahmad, R},
  booktitle={4th International Conference on Mechanical Engineering},
  pages={31--36},
  year={2001}
}

@article{Devinant2002,
  title={Experimental study of wind-turbine airfoil aerodynamics in high turbulence},
  author={Devinant, Ph and Laverne, Thomas and Hureau, Jacques},
  journal={Journal of Wind Engineering and Industrial Aerodynamics},
  volume={90},
  number={6},
  pages={689--707},
  year={2002},
  publisher={Elsevier},
  doi={10.1016/S0167-6105(02)00162-9}
}

@article{Kay2020,
  title={Influence of Turbulence on Cambered and Symmetrical Airfoils at Low Reynolds Numbers},
  author={Kay, Nicholas J and Richards, Peter J and Sharma, Rajnish N},
  journal={AIAA journal},
  volume={58},
  number={5},
  pages={1913--1925},
  year={2020},
  publisher={American Institute of Aeronautics and Astronautics},
  doi={10.2514/1.J058822}
}

@article{Damiola2023,
  title={Influence of free-stream turbulence intensity on static and dynamic stall of a NACA 0018 aerofoil},
  author={Damiola, Luca and Siddiqui, Muhammad Faheem and Runacres, Mark Charles and De Troyer, Tim},
  journal={Journal of Wind Engineering and Industrial Aerodynamics},
  volume={232},
  pages={105270},
  year={2023},
  publisher={Elsevier},
  doi={10.1016/j.jweia.2022.105270}
}

@article{Vita2020,
  title={The effect of the integral length scale of turbulence on a wind turbine aerofoil},
  author={Vita, Giulio and Hemida, Hassan and Andrianne, Thomas and Baniotopoulos, Charalampos},
  journal={Journal of Wind Engineering and Industrial Aerodynamics},
  volume={204},
  pages={104235},
  year={2020},
  publisher={Elsevier},
  doi={10.1016/j.jweia.2020.104235}
}

@article{Thompson2023,
  title={Effects of integral length scale variations on the stall characteristics of a wing at high free-stream turbulence conditions},
  author={Thompson, C and Biler, H and Symon, S and Ganapathisubramani, B},
  journal={Journal of Fluid Mechanics},
  volume={974},
  pages={A9},
  year={2023},
  publisher={Cambridge University Press},
  doi={10.1017/jfm.2023.789}
}

@article{Wang2014,
  title     = {Turbulent intensity and {Reynolds} number effects on an airfoil at low {Reynolds} numbers},
  author    = {Wang, S. and Zhou, Y. and Alam, Md. Mahbub and Yang, Hongxing},
  journal   = {Physics of Fluids},
  volume    = {26},
  number    = {11},
  pages     = {115107},
  year      = {2014},
  publisher = {AIP Publishing},
  doi       = {10.1063/1.4901969}
}

@article{Li2021,
  title={The influence of freestream turbulence on the temporal pressure distribution and lift of an airfoil},
  author={Li, Leon and Hearst, R Jason},
  journal={Journal of Wind Engineering and Industrial Aerodynamics},
  volume={209},
  pages={104456},
  year={2021},
  publisher={Elsevier},
  doi={10.1016/j.jweia.2020.104456}
}

@article{Istvan2018,
  title={Effects of free-stream turbulence intensity on transition in a laminar separation bubble formed over an airfoil},
  author={Istvan, Mark S and Yarusevych, Serhiy},
  journal={Experiments in Fluids},
  volume={59},
  number={3},
  pages={52},
  year={2018},
  publisher={Springer},
  doi={10.1007/s00348-018-2511-6}
}

@article{Zaman1989,
  title={A natural low-frequency oscillation of the flow over an airfoil near stalling conditions},
  author={Zaman, KBMQ and McKinzie, DJ and Rumsey, CL},
  journal={Journal of Fluid Mechanics},
  volume={202},
  pages={403--442},
  year={1989},
  publisher={Cambridge University Press},
  doi={10.1017/S0022112089001230}
}

@phdthesis{Sheppard2025,
author={Sheppard,Samantha J.},
year={2025},
title={Generalizing the Surface Layer: Investigating the Physical Mechanisms by Which Impermeable Surfaces Alter Turbulence Structure},
journal={ProQuest Dissertations and Theses},
pages={234},
isbn={9798291576632},
school={University of Colorado Boulder}
}

@article{Roach1987,
title = {The generation of nearly isotropic turbulence by means of grids},
journal = {International Journal of Heat and Fluid Flow},
volume = {8},
number = {2},
pages = {82-92},
year = {1987},
doi = {10.1016/0142-727X(87)90001-4},
author = {P.E. Roach},
}

@article{Amitay2002,
author = {Amitay, Michael and Glezer, Ari},
title = {Role of Actuation Frequency in Controlled Flow Reattachment over a Stalled Airfoil},
journal = {AIAA Journal},
volume = {40},
number = {2},
pages = {209-216},
year = {2002},
doi = {10.2514/2.1662},
}

@article{Hurst2007,
    author = {Hurst, D. and Vassilicos, J. C.},
    title = {Scalings and decay of fractal-generated turbulence},
    journal = {Physics of Fluids},
    volume = {19},
    number = {3},
    pages = {035103},
    year = {2007},
    month = {03},
    doi = {10.1063/1.2676448},
}

@article{Nikoueeyan2023,
    author = {Nikoueeyan, P. and Naughton, J.W.},
    title = {Analysis of the unsteady surface pressure distribution of a pitching airfoil using modal decomposition},
    journal = {Experiments in Fluids},
    year = {2023},
number = {64},
issue = {100},
doi = {10.1007/s00348-023-03644-5},
}

@article{Taira2017,
author = {Taira, Kunihiko and Brunton, Steven L. and Dawson, Scott T. M. and Rowley, Clarence W. and Colonius, Tim and McKeon, Beverley J. and Schmidt, Oliver T. and Gordeyev, Stanislav and Theofilis, Vassilios and Ukeiley, Lawrence S.},
title = {Modal Analysis of Fluid Flows: An Overview},
journal = {AIAA Journal},
volume = {55},
number = {12},
pages = {4013-4041},
year = {2017},
doi = {10.2514/1.J056060},
}

@article{Taira2020,
author = {Taira, Kunihiko and Hemati, Maziar S. and Brunton, Steven L. and Sun, Yiyang and Duraisamy, Karthik and Bagheri, Shervin and Dawson, Scott T. M. and Yeh, Chi-An},
title = {Modal Analysis of Fluid Flows: Applications and Outlook},
journal = {AIAA Journal},
volume = {58},
number = {3},
pages = {998-1022},
year = {2020},
doi = {10.2514/1.J058462},
}

@article{Hristov2018,
author = {Hristov, Georgi and Ansell, Phillip J.},
title = {Poststall Hysteresis and Flowfield Unsteadiness on a NACA 0012 Airfoil},
journal = {AIAA Journal},
volume = {56},
number = {7},
pages = {2528-2539},
year = {2018},
doi = {10.2514/1.J056774},
}

@article{Mueller1985,
author = {Mueller, Thomas J.},
title = {The influence of laminar separation and transition on low Reynolds number airfoil hysteresis},
journal = {Journal of Aircraft},
volume = {22},
number = {9},
pages = {763-770},
year = {1985},
doi = {10.2514/3.45199},
}

@article{Mishra2022,
  author  = {Mishra, Rishabh and Neunaber, Ingrid and Guilmineau, Emmanuel and Braud, Caroline},
  title   = {Wind tunnel study: is turbulent intensity a good candidate to help in bypassing low {R}eynolds number effects on 2{D} blade sections?},
  journal = {Journal of Physics: Conference Series},
  volume  = {2265},
  number  = {2},
  pages   = {022095},
  year    = {2022},
  doi     = {10.1088/1742-6596/2265/2/022095}
}

@article{Sicot2006,
  title={Unsteady characteristics of the static stall of an airfoil subjected to freestream turbulence level up to 16\%},
  author={Sicot, Christophe and Aubrun, Sandrine and Loyer, Stephane and Devinant, Philippe},
  journal={Experiments in fluids},
  volume={41},
  number={4},
  pages={641--648},
  year={2006},
  publisher={Springer},
  doi={10.1007/s00348-006-0187-9}
}

@article{Busquet2021,
  title={Bifurcation scenario for a two-dimensional static airfoil exhibiting trailing edge stall},
  author={Busquet, Denis and Marquet, Olivier and Richez, Fran{\c{c}}ois and Juniper, Matthew and Sipp, Denis},
  journal={Journal of Fluid Mechanics},
  volume={928},
  pages={A3},
  year={2021},
  publisher={Cambridge University Press},
  doi={doi.org/10.1017/jfm.2021.767}
}

@article{Neunaber2022,
  title={Wind tunnel study on natural instability of the normal force on a full-scale wind turbine blade section at Reynolds number $4.7\cdot 10^6$},
  author={Neunaber, Ingrid and Danbon, Fr{\'e}d{\'e}ric and Soulier, Antoine and Voisin, Dimitri and Guilmineau, Emmanuel and Delpech, Philippe and Courtine, S{\'e}bastien and Taymans, Claire and Braud, Caroline},
  journal={Wind Energy},
  volume={25},
  number={8},
  pages={1332--1342},
  year={2022},
  publisher={Wiley Online Library},
  doi={10.1002/we.2732}
}

@article{Braud2024,
  title={Study of the wall pressure variations on the stall inception of a thick cambered profile at high Reynolds number},
  author={Braud, Caroline and Podvin, B{\'e}reng{\`e}re and Deparday, Julien},
  journal={Physical Review Fluids},
  volume={9},
  number={1},
  pages={014605},
  year={2024},
  publisher={APS},
  doi={10.1103/PhysRevFluids.9.014605}
}

@article{Hanna2026,
  title={Tracking stall cell dynamics at high Reynolds numbers},
  author={Hanna, Badoui and Podvin, B{\'e}reng{\`e}re and Braud, Caroline},
  journal={arXiv Preprint},
  year={2026},
  doi={10.48550/arXiv.2602.03844}
}

@article{Veers2019,
  author  = {Veers, Paul and Dykes, Katherine and Lantz, Eric and Barth, Stephan and Bottasso, Carlo L. and Carlson, Ola and Clifton, Andrew and Green, Johney and Green, Peter and Holttinen, Hannele and others},
  title   = {Grand challenges in the science of wind energy},
  journal = {Science},
  volume  = {366},
  number  = {6464},
  pages   = {eaau2027},
  year    = {2019},
  doi     = {10.1126/science.aau2027}
}

\end{document}